\documentclass[useAMS,usenatbib,a4paper]{mn2e}
\usepackage{graphicx}
\usepackage{aas_macros}
\usepackage{multirow}
\usepackage{hyperref}
\usepackage{subfigure}
\title[Nitrogen Enrichment in Wolf-Rayet Ejecta Nebulae]{{\em Herschel\thanks{{\it Herschel} is an ESA space observatory with science instruments provided by European-led Principal Investigator consortia and with important participation from NASA.}}-PACS Measurements of Nitrogen Enrichment in Nebulae around Wolf-Rayet Stars} 

\author[D. J. Stock \& M. J. Barlow]{D. J. Stock$^{1}$\thanks{E-mail: 
dstock4@uwo.ca (DS)} \& M.J. Barlow$^{2}$\\ 
$^{1}$ Department of Physics and Astronomy, University of Western Ontario, 
Richmond Street, London, ON, N6A 3K7, Canada\\
$^{2}$ Department of Physics and Astronomy, University College London, Gower 
Street, London, WC1E 6BT, UK\\
}

\begin{document}

\date{MNRAS accepted: 9 April 2014}

\pagerange{\pageref{firstpage}--\pageref{lastpage}} \pubyear{2014}

\maketitle

\label{firstpage}

\begin{abstract} For three nebulae that have early-WN Wolf-Rayet exciting 
stars, NGC~6888, WR~8 and Abell~48, we have obtained {\em Herschel}-PACS line 
scans of the [N~{\sc iii}] 57 \micron\ and [O~{\sc iii}] 88 \micron\ 
lines, along with the 122 and 205 \micron\ lines of [N~{\sc ii}]. From the 
former two lines we have derived N$^{2+}$/O$^{2+}$ abundance ratios, equal 
to the overall N/O ratio under a wide range of nebular conditions. We find 
that all of the nebulae observed possess significant nitrogen enrichment, 
with derived N/O ratios greater than solar. The two nebulae with massive 
Wolf-Rayet exciting stars, NGC~6888 and WR~8 are found to have N/O ratios 
that are enhanced by factors of 7 - 10 relative to the solar N/O ratio, 
consistent with an origin as material ejected just before the onset of the 
Wolf-Rayet phase. The other nebula, Abell~48, has recently been 
reclassified as a member of the rare class of three planetary nebulae that have 
early-WN central stars and are not of Peimbert Type I. We derive a nebular N/O ratio for it that is a 
factor of 4 enhanced relative to solar and slightly above the range of N/O 
values that have been measured for the other three members of its [WN] 
planetary nebula class.

\end{abstract}

\begin{keywords}
stars: Wolf-Rayet, ISM: abundances, stars: mass-loss, stars: massive, stars: winds, outflows, circumstellar matter
\end{keywords}

\section{Introduction}

Core-collapse supernovae from massive stars are known to dominate the 
production of oxygen and heavier $\alpha$-particle elements in galaxies 
but massive stars can make an important contribution to the nitrogen 
enrichment of the ISM during the early evolution of galaxies 
\citep{2000ApJ...541..660H}. This can happen via the ejection of 
nitrogen-enriched material that has been heavily processed by the 
CNO-cycle during their Luminous Blue Variable (LBV) and ensuing WN 
Wolf-Rayet (WR) stages.

Observational determinations of the quantities of nitrogen injected into 
the ISM by WN stars and/or their LBV precursors are limited in number. However, a subset of WR 
stars are surrounded by nebulae whose expansion ages, morphologies, and 
occasional abundance determinations (e.g. \citealt{1992AA...259..629E}; 
\citealt{2011MNRAS.418.2532S}), show them to consist of nitrogen-rich 
material that has been recently ejected by the star; these are termed 
`ejecta nebulae' (\citealt{1991IAUS..143..349C}; 
\citealt{2010MNRAS.409.1429S}).

Amongst the small sample of known, or candidate, Wolf-Rayet ejecta 
nebulae, only a fraction have been probed spectroscopically and their 
abundances derived. Such efforts have been hampered by the high degree of 
extinction along their sight lines in the galactic plane. The usual 
suite of abundance determination methods (e.g. 
\citealt{1977RMxAA...2..181T}; \citealt{1994MNRAS.271..257K}; 
\citealt{2012MNRAS.422.3516W}) rely on complex empirical relationships to 
calibrate the densities, temperatures and ionization schemes used to 
derive reliable abundances. The main stumbling block for these schemes 
with respect to nebulae around hot, massive stars comes in the lack of 
optical transitions from some of the ionization states that are expected 
to dominate the ion populations. For example, nitrogen abundances in 
nebulae are commonly calculated using the [N~{\sc ii}] 6548/6584 \AA\ 
doublet, however N$^+$ is often not the dominant nitrogen ion, so 
ionization correction factors are employed to attempt to compensate. In 
addition heavy extinction by interstellar dust towards galactic plane WR 
stars typically means that the blue end of the optical spectrum is 
missing, and therefore abundance determinations cannot be easily made 
(e.g. \citealt{2011MNRAS.418.2532S}). At far-IR wavelengths the effect of 
dust extinction is essentially removed and the usually dominant ionization 
stages of oxygen and nitrogen (N$^{2+}$ and O$^{2+}$) have accessible fine 
structure lines whose low excitation energies render ionic abundances 
derived from them insensitive to the adopted nebular electron temperature.

Using the PACS instrument \citep{2010A&A2...518L...2P} on the {\em Herschel Space Observatory} \citep{2010A&A...518L...1P}, we have obtained 
spectral data cubes of various fine structure lines in the PACS wavelength 
range. For each source the [N~{\sc iii}] and [O~{\sc iii}] lines at 57 and 
88 \micron\ have been mapped, along with the [N~{\sc ii}] lines at 122 and 
205 \micron, which we will employ to test assumptions about a) density 
and b) the ionization structure.

Our {\em Herschel} program (\verb+OT2_dstock_3+) covered two WR ejecta 
nebulae (around WR~8 and WR~136, the latter's nebula better known as NGC 
6888) along with the PN A66 48 (henceforth Abell~48). At the time 
of its PACS observations in 2012, Abell~48 was thought to have been generated by 
a massive WR star but two recent studies have shown it to be a classical 
PN, albeit one with a rare WN-type central star 
(\citealt{2013MNRAS.430.2302T2}; \citealt{2013arXiv1301.3994F}).
 
\section{Data}

\subsection{Target Selection}

Our target sample of was drawn from the Wolf-Rayet ejecta nebula catalogue 
of \citet{2010MNRAS.409.1429S}. A selection criterion was that the 
exciting WR star should have a spectral type of WN7 or earlier, or WC7 or 
earlier, in order that N$^{2+}$ and O$^{2+}$ would be the dominant ion 
stages of N and O, respectively (this is unlikely to be the case for 
later-type, lower-excitation WR central stars).

The ejecta nebula around WR~8 has been observed spectroscopically in the 
optical \citep{2011MNRAS.418.2532S} but due to heavy reddening nebular 
temperature diagnostics were not available and so no reliable abundances 
could be derived. However the lower limits quoted by 
\citet{2011MNRAS.418.2532S} suggest that the nebula around WR~8 is heavily 
enriched in nitrogen. The spectral type of WR~8 (WC4/WN7) indicates that 
it is a transition star, believed to be changing from its nitrogen rich 
phase (WN) to its carbon rich phase (WC) \citep{C95}. As such the total 
amount of nitrogen in the nebula should be close to the total amount of 
nitrogen that will ever be ejected by this star.

The second target, NGC 6888, is one of the best known WR nebulae, having 
been studied in great detail by many authors (most recently by 
\citet{2012AA...541A.119F} who present a thorough review). It is thought 
to be the result of a major outflow of material whilst the parent star (WR 
136; HD 192163; spectral type WN6h) was passing through a red giant phase 
\citep{M95}.

The third target, the nebula PN Abell~48, has been found to 
possess an early-WN type central star \citep{2010AJ....139.2330W}. More 
recent work (\citealt{2013MNRAS.430.2302T2}; 
\citealt{2013arXiv1301.3994F}) has shown that this object is actually much 
more likely to be a rare type of planetary nebula (PN), with a nitrogen 
rich low mass central star (of type [WN]) which mimics the spectral 
appearance of a more massive WN star. The exact subtype of the central 
star has been debated, with the \citet{2013MNRAS.430.2302T2} study 
estimating it to be [WN~5] while \citet{2013arXiv1301.3994F} estimated it to be [WN~4--5]. The arguments presented by 
\citet{2013arXiv1301.3994F} to show that the central star is of low mass 
are twofold. Firstly, if it were a massive star it would be at a very 
large distance and the mass of the expelled material would be around 50 
M$_{\sun}$, which is above the initial mass range for massive stars which 
are thought to become mid-WN subtypes \citep{C07}. Secondly, the 
increased distance would place the system behind the galactic bulge and 
therefore change the expected extinction dramatically, which is not 
confirmed by their spectra.

\subsection{Observations}


\begin{table*}
	\begin{center}
	\caption{Log of Observations}	
	\label{tab:obs}

	\subfigure[General Characteristics]{
	\begin{tabular}{lccp{0.9cm}ccp{2.25cm}p{2.3cm}}
	\hline	
	Object          & RA$^a$ & Dec$^b$ & Chop Throw & ObsID & Date and Time$^c$ & Lines & Total Observing Time [s]$^d$ \\
	\hline
	WR 8            & 07 44 50.83 & -31 55 15.8 & - & 1342245246 & 2012-05-02 10:35:04 & [O~{\sc iii}] 88 \micron, [N~{\sc ii}] 205 \micron & 4646\\
	WR 8 sky        & 07 44 17.68 & -31 52 10.6 & - & 1342245246 & 2012-05-02 10:35:04 & [O~{\sc iii}] 88 \micron, [N~{\sc ii}] 205 \micron & \\
	WR 8            & 07 44 50.83 & -31 55 15.8 & - & 1342245247 & 2012-05-02 11:54:37 & [N~{\sc iii}] 57 \micron, [N~{\sc ii}] 122 \micron & 1577\\
	WR 8 sky        & 07 44 17.68 & -31 52 10.6 & - & 1342245247 & 2012-05-02 11:54:37 & [N~{\sc iii}] 57 \micron, [N~{\sc ii}] 122 \micron & \\
	A 48            & 18 42 46.92 & -3 13 17.2 & 1\arcmin & 1342253738 & 2012-10-21 03:04:59 & [N~{\sc iii}] 57 \micron, [N~{\sc ii}] 122, 205 \micron & 2371\\
	A 48            & 18 42 46.92 & -3 13 17.2 & 1\arcmin & 1342253739 & 2012-10-21 03:46:43 & [O~{\sc iii}] 88 \micron & 814\\
	NGC 6888 (rim)  & 20 12 39.02 & 38 26 55.1 & - & 1342256477 & 2012-12-06 13:36:56 & [O~{\sc iii}] 88 \micron, [N~{\sc ii}] 205 \micron & 2319\\
	NGC 6888 sky    & 20 11 34.89 & 38 25 19.9 & - & 1342256477 & 2012-12-06 13:36:56 & [O~{\sc iii}] 88 \micron, [N~{\sc ii}] 205 \micron & \\
	NGC 6888 (rim)  & 20 12 39.02 & 38 26 55.1 & - & 1342256478 & 2012-12-06 14:17:45 & [N~{\sc iii}] 57 \micron, [N~{\sc ii}] 122 \micron & 1570\\
	NGC 6888 sky    & 20 11 34.89 & 38 25 19.9 & - & 1342256478 & 2012-12-06 14:17:45 & [N~{\sc iii}] 57 \micron, [N~{\sc ii}] 122 \micron & \\
	NGC 6888 (inner)& 20 11 56.02 & 38 21 40.1 & - & 1342256927 & 2012-12-09 15:51:16 & [O~{\sc iii}] 88 \micron, [N~{\sc ii}] 205 \micron & 1574\\
	NGC 6888 sky    & 20 11 34.89 & 38 25 19.9 & - & 1342256927 & 2012-12-09 15:51:16 & [O~{\sc iii}] 88 \micron, [N~{\sc ii}] 205 \micron & \\
	NGC 6888 (inner)& 20 11 56.02 & 38 21 40.1 & - & 1342256928 & 2012-12-09 16:19:46 & [N~{\sc iii}] 57 \micron, [N~{\sc ii}] 122 \micron & 2323\\
	NGC 6888 sky    & 20 11 34.89 & 38 25 19.9 & - & 1342256928 & 2012-12-09 16:19:46 & [N~{\sc iii}] 57 \micron, [N~{\sc ii}] 122 \micron & \\
	\end{tabular}
	\label{tab:obs1}
	}
	\vspace{0.5cm}
	\subfigure[Exposure Times and Observation Characteristics]{
	\begin{tabular}{lccccp{1.5cm}}
	\hline	
	Object          & Line & Wavelength Range [\micron]&  Resolving Power & Repetitions & Exposure Time [s] \\
	\hline
	WR 8            & [N~{\sc iii}] 57 \micron & 56.95 -- 57.72   & $\simeq$ 2800 & 3 & 225 \\
	WR 8            & [O~{\sc iii}] 88 \micron & 87.77 -- 88.98   & $\simeq$ 2400 & 2 & 300 \\
	WR 8            & [N~{\sc ii}] 122 \micron & 119.98 -- 123.77 & $\simeq$ 1100 & 3 & 225 \\
	WR 8            & [N~{\sc ii}] 205 \micron & 204.11 -- 206.19 & $\simeq$ 1700 & 8 & 1200\\
	Abell~48            & [N~{\sc iii}] 57 \micron & 57.05 -- 57.64   & $\simeq$ 2800 & 2 & 384 \\
	Abell~48            & [O~{\sc iii}] 88 \micron & 87.92 -- 88.81   & $\simeq$ 2400 & 2 & 368 \\
	Abell~48            & [N~{\sc ii}] 122 \micron & 120.57 -- 123.18 & $\simeq$ 1100 & 2 & 344 \\
	Abell~48            & [N~{\sc ii}] 205 \micron & 204.58 -- 206.04 & $\simeq$ 1700 & 2 & 344 \\
	NGC 6888 (rim)  & [N~{\sc iii}] 57 \micron & 56.95 -- 57.73   & $\simeq$ 2800 & 3 & 225 \\
	NGC 6888 (rim)  & [O~{\sc iii}] 88 \micron & 87.77 -- 88.98   & $\simeq$ 2400 & 2 & 150 \\
	NGC 6888 (rim)  & [N~{\sc ii}] 122 \micron & 119.98 -- 123.77 & $\simeq$ 1100 & 3 & 225 \\
	NGC 6888 (rim)  & [N~{\sc ii}] 205 \micron & 204.16 -- 206.24 & $\simeq$ 1700 & 8 & 600 \\
	NGC 6888 (inner)& [N~{\sc iii}] 57 \micron & 56.95 -- 57.73   & $\simeq$ 2800 & 3 & 225 \\
	NGC 6888 (inner)& [O~{\sc iii}] 88 \micron & 87.77 -- 88.98   & $\simeq$ 2400 & 2 & 150 \\
	NGC 6888 (inner)& [N~{\sc ii}] 122 \micron & 119.98 -- 123.77 & $\simeq$ 1100 & 3 & 225 \\
	NGC 6888 (inner)& [N~{\sc ii}] 205 \micron & 204.16 -- 206.24 & $\simeq$ 1700 & 8 & 600 \\
	\end{tabular}
	\label{tab:obs2}
	}
	
	\end{center}	

	\bigskip
	$^a$: Hours, minutes and seconds. $^b$: Degrees, arcminutes and arcseconds. $^c$: UT. $^d$: for WR 8 and NGC 6888 the sky positions are included in the total OB execution time.
\end{table*}

\begin{figure*}
	\centering
	\includegraphics[width=170mm]{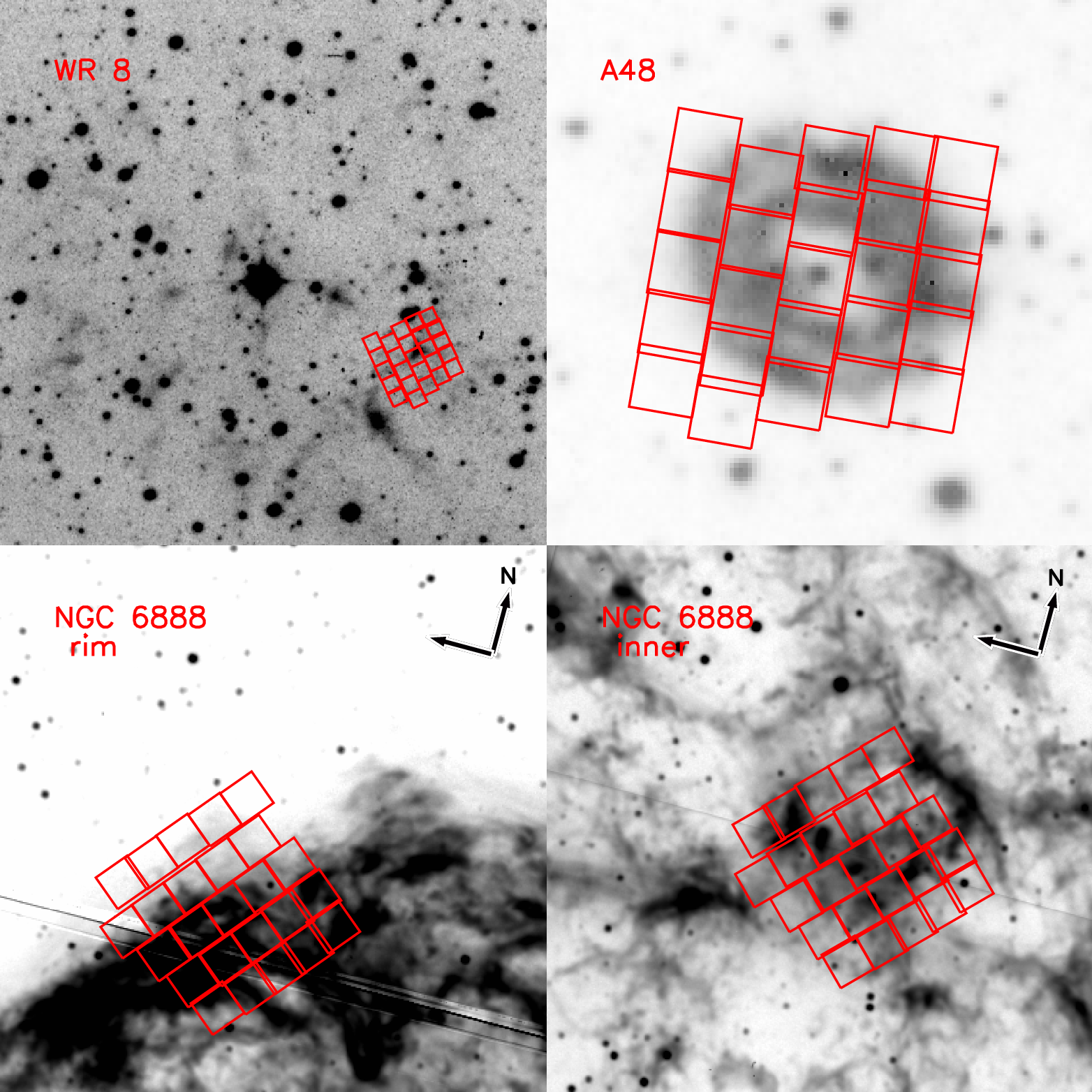}	
 	\caption{Herschel PACS spectrometer fields of view for observed 
sources. The array of red boxes represent the footprint of the 5$\times$5 
array of 9.4\arcsec square pixels. North is up and east is to the left 
except for the NGC 6888 fields where its direction is indicated. 
Background images are Supercosmos H$\alpha$ Survey imagery for WR~8 and Abell~48 \citep{P05} and 
IPHAS survey images for NGC 6888 \citep{D05}. }
	\label{fig:fovs}
\end{figure*}

Herschel/PACS spectra of four fine structure lines between $\sim$55 and 
$\sim$205 \micron\ were obtained in line scan mode for each of the objects 
discussed in the previous section. The PACS spectrometer is an integral 
field unit (IFU) consisting of a 5~$\times$~5 array of 
9.4\arcsec~$\times$~9.4\arcsec\ spectral pixels (spaxels). Details of the 
observations are presented in Table~\ref{tab:obs}, with general information 
in Table~\ref{tab:obs1} and line scan details, including exposure times, in 
Table~\ref{tab:obs2}. Figure~\ref{fig:fovs} shows the spectrometer IFU
footprint superimposed on Supercosmos H$\alpha$ images of the targets.

The PACS spectrometer chopping nodding observation mode was employed for 
the smallest nebula (Abell~48), while for the larger nebulae the chop throw 
was not sufficient to find a `sky' field. For WR~8 and NGC~6888 we 
therefore used the pointed line scan mode, using specific sky fields far 
from the nebulae. In both types of observation the `sky' fields were used 
to cancel the emission from the telescope and instruments.

For Abell~48 and WR~8, one on-source field was observed for each object. In 
fact Abell~48 nearly fills the entire PACS IFU spectrometer field of view, while for 
the nebula around WR~8 we selected the brightest H$\alpha$ patch of 
nebulosity. The large angular size of NGC~6888 allowed us to observe two fields, 
one at the rim and and one near the centre.

Each line scan observation resulted in a spectral cube containing two 
spatial dimensions with five spaxels along each axis and a wavelength axis 
encompassing the spectral line and a section of continuum on either side. 
The spectral resolution of the PACS instrument varies with wavelength and 
grating order. For [N~{\sc iii}] 57 \micron, R $\sim$ 2700, while for the 
[N~{\sc ii}] line at 122 \micron\ the resolving power was 
1100\footnotemark. The wavelength ranges and resolving powers are given for 
each observation in Table~\ref{tab:obs2}.

\footnotetext{PACS Observers Manual: \\ 
\url{http://herschel.esac.esa.int/Docs/PACS/pdf/pacs_om.pdf}}

\subsection{Data Reduction and Analysis}

Each of the observations was reduced using the standard PACS spectroscopy 
reduction pipeline within HIPE (v11.0.1; \citealt{2010ASPC..434..139O2}).

The analysis process for the data was relatively simple. Initially, the 
alignments for each of the different data cubes were checked for each 
object, i.e. whether the observations of each line covered precisely the 
same area. It was found that in all cases the deviations were less than 1 
\arcsec due to slight rotation of the telescope between observations, 
relative to the field. In order to minimise the effect of the slightly 
mis-matched apertures on the later results, the integrated spectra were 
calculated using in inner 3$\times$3 array of pixels rather than the whole 
5$\times$5 array.

\begin{figure*}
	\centering
	\includegraphics[width=130mm]{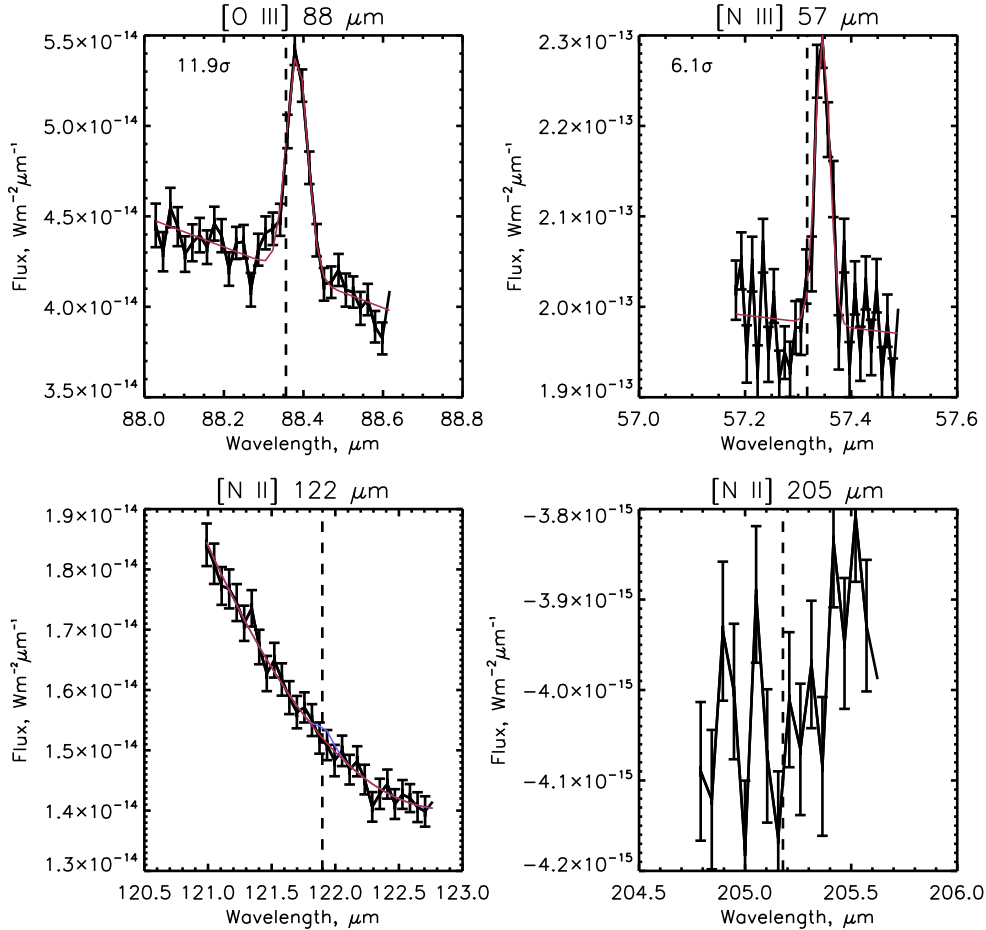}	
	\caption{Integrated spectra (black) and line fits (red) to PACS line scans 
for the nebula around WR~8, clockwise from top left: [O~{\sc iii}] 88 
\micron, [N~{\sc iii}] 57 \micron, [N~{\sc ii}] 205 \micron, [N~{\sc ii}] 
122 \micron. The derived line signal to noise ratios are indicated for each 
fit. For the [N~{\sc ii}] 122 \micron\ line we include a gaussian with the upper flux limit (blue).}
	\label{fig:fitswr8}
\end{figure*}

\begin{figure*}
	\centering
	\includegraphics[width=130mm]{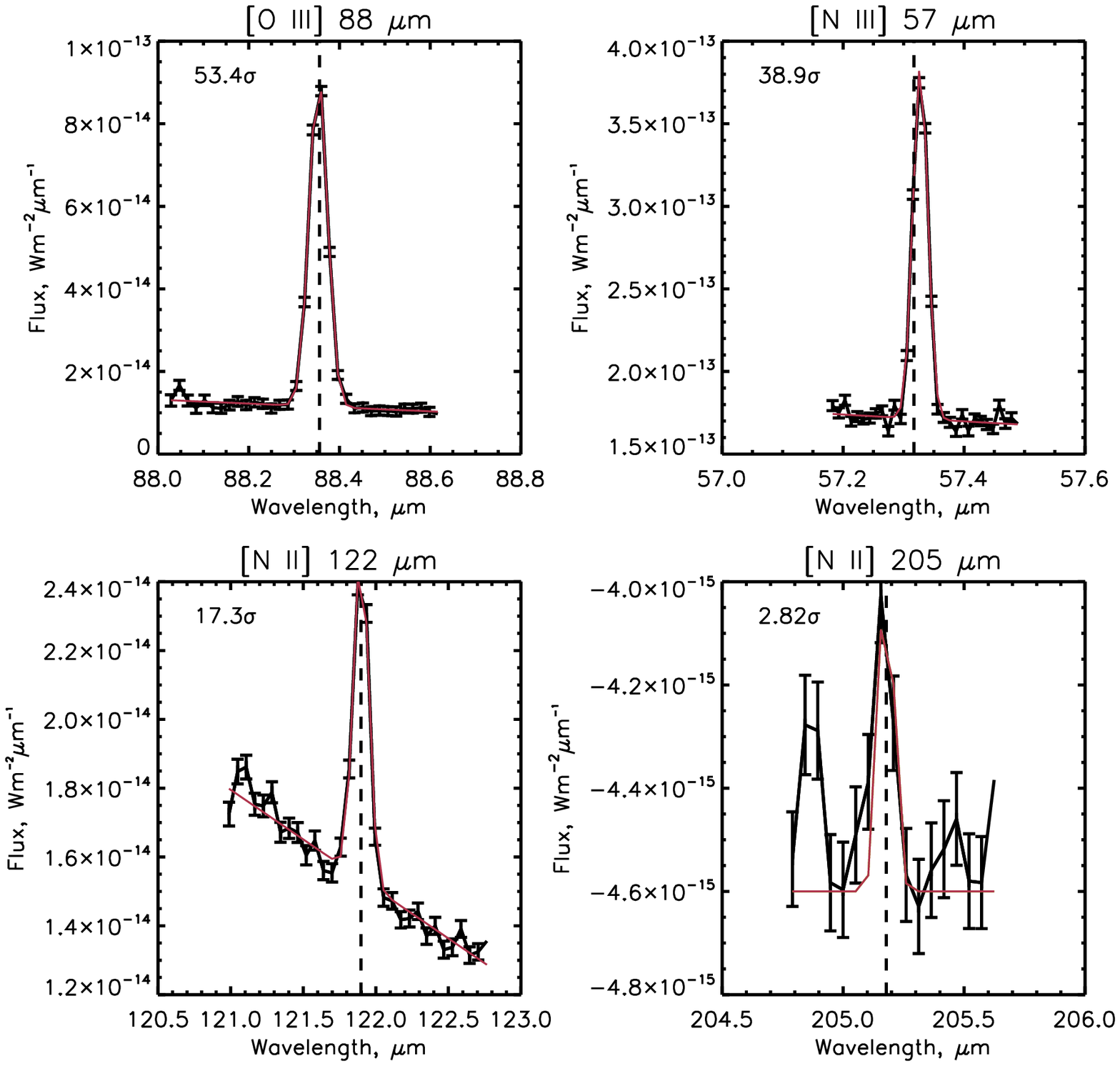}	
	\caption{Integrated spectra (black) and line fits (red) to PACS line scans for 
NGC 6888 (outer), clockwise from top left: [O~{\sc iii}] 88 \micron, 
[N~{\sc iii}] 57 \micron, [N~{\sc ii}] 205 \micron, [N~{\sc ii}] 122 \micron. 
Line signal to noise ratios are indicated for each fit.}
	\label{fig:fits68881}
\end{figure*}

\begin{figure*}
	\centering
	\includegraphics[width=130mm]{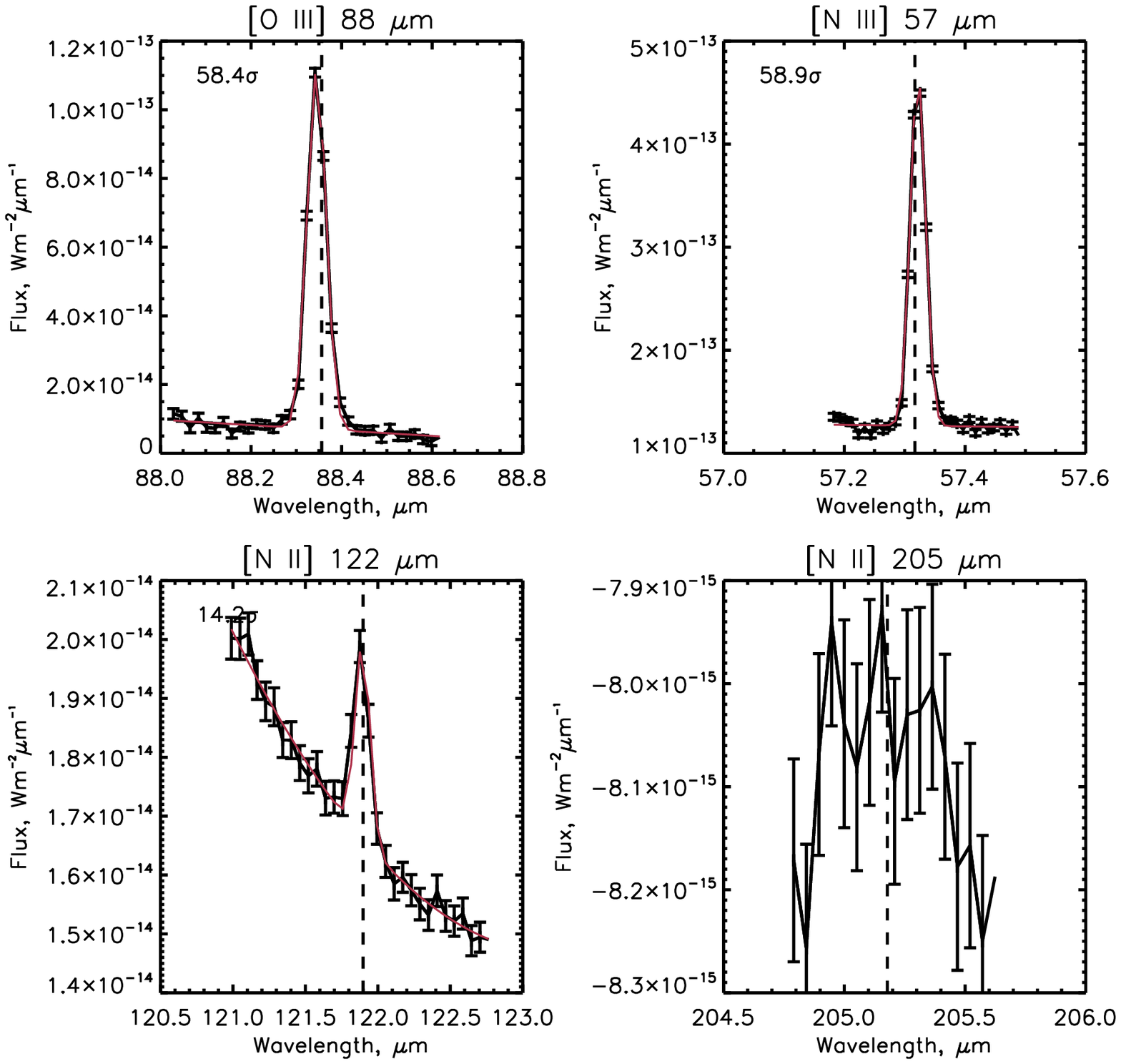}	
	\caption{Integrated spectra (black) and line fits (red) to PACS line scans for 
NGC 6888 (inner), clockwise from top left: [O~{\sc iii}] 88 \micron, 
[N~{\sc iii}] 
57 \micron, [N~{\sc ii}] 205 \micron, [N~{\sc ii}] 122 \micron. Line signal 
to noise ratios are indicated for each fit.}
	\label{fig:fits68882}
\end{figure*}

\begin{figure*}
	\centering
	\includegraphics[width=130mm]{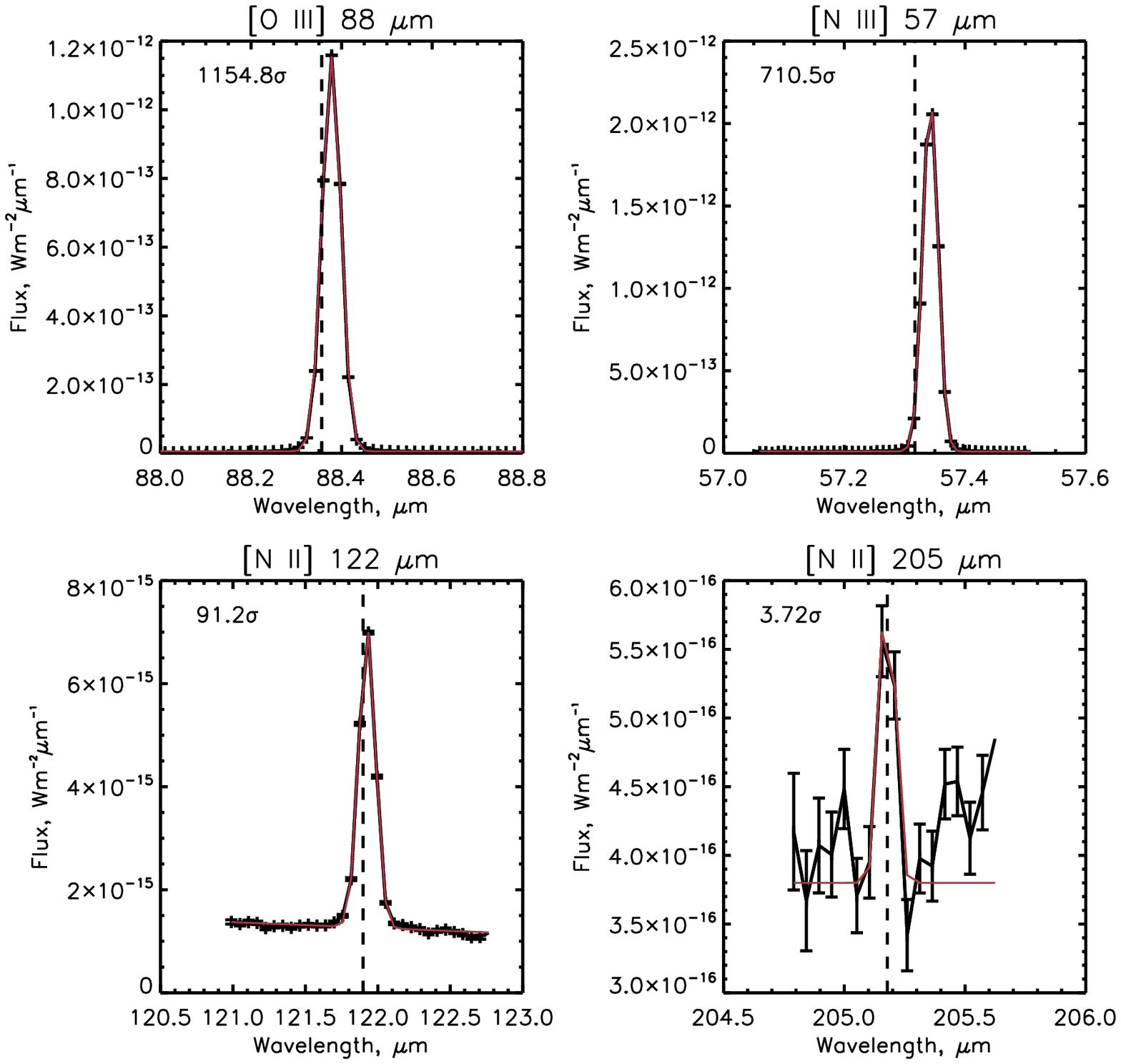}	
	\caption{Integrated spectra (black) and line fits (red) to PACS line scans for 
Abell~48, clockwise from top left: [O~{\sc iii}] 88 \micron, [N~{\sc iii}] 57 
\micron, [N~{\sc ii}] 205 \micron, [N~{\sc ii}] 122 \micron. Line signal 
to noise ratios are indicated for each fit.}
	\label{fig:fitsa48}
\end{figure*}

Subsequently, lines in the four spectral datacubes for each observation 
were measured identically. Each coadded spectrum was fitted with a straight 
line plus a gaussian, representing the continuum and line emission 
respectively (see Figures~\ref{fig:fitswr8}--\ref{fig:fitsa48} for an 
example). The continuum was initially fitted to `windows' of 10 or so 
pixels on either side of the line to gain a rough estimate of the continuum 
properties, which were then used as initial conditions for the fit to the 
continuum and line. For the two instances of [N~{\sc ii}] 205 \micron\ line 
detections a slightly different fitting routine was used because the 
increased noise in this region rendered the method unreliable. Instead a 
gaussian was fitted to the data assuming a continuum, which was set to be 
constant at the level of the base of the detected line. 

A further exception was made for the fit to the weak [N~{\sc ii}] 122 
\micron\ line from the inner region of NGC~6888. For this spectrum (see 
Figure~\ref{fig:fits68882}) there is clear curvature to the continuum and 
so the straight line continuum fit was modified to a second order polynomial for this 
source.

An upper limit to the [N~{\sc ii}] 122 \micron\ flux for WR 8 was found by initially fitting
 a second order polynomial continuum to the observation. Subsequently a gaussian with the same width as 
that found for A48 was added and the flux tuned such that the line could just be clearly discerned against the noise - this was adopted
as the 3$\sigma$ upper limit to the line flux.

In order to calculate the signal to noise ratio (SNR), we measured the root 
mean square (RMS) noise of the continuum measurement by subtracting the 
fitted continuum from the data and calculating the RMS within the spectral
windows previously mentioned. The SNR of each line was estimated by 
comparing the integrated line flux to the RMS noise taking into account 
the width of the gaussian and wavelength spacing of the spectral points.

\section{Results}

In Table~\ref{tab:intline} we list integrated line fluxes and some 
derived quantities for the central 3$\times$3 group of spaxels for the 
lines of [N~{\sc iii}] 57 \micron, [O~{\sc iii}] 88 \micron\ and [N~{\sc 
ii}] 122, 205 \micron. Figures with the line fits to the integrated 
spectra are shown for WR~8 in Figure~\ref{fig:fitswr8}, the outer field of 
NGC 6888 in Figure~\ref{fig:fits68881}, the inner field of NGC 6888 in 
Figure~\ref{fig:fits68882} and for Abell~48 in Figure~\ref{fig:fitsa48}.

\begin{table*}
	\begin{center}
	\caption{Integrated Nebular Properties}\label{tab:intline}
	\begin{tabular}{lccccccc}
	\hline
		& \multicolumn{4}{c}{Fluxes$^a$}&\multicolumn{2}{c}{Abundances}\\	
	Object  & [O~{\sc iii}] 88~$\mu$m & [N~{\sc iii}] 57~$\mu$m & [N~{\sc ii}] 122 \micron & [N~{\sc ii}] 205 \micron & N$^{2+}$/N$^{+}$ & N$^{2+}$/O$^{2+}$=N/O & n$_e$(cm$^{-3}$)\\
	\hline
WR 8             & 8.08 $\pm$ 0.68 & 11.11 $\pm$ 1.81 & $<$0.30 & -- & $>$7.78 & 0.92 $\pm$ 0.17 & 50$^b$\\
NGC 6888 (rim)   & 40.34 $\pm$ 0.76 & 65.88 $\pm$ 1.70 & 12.55 $\pm$ 0.72 & 0.50 $\pm$ 0.18 & 0.68 $\pm$ 0.07 & 0.93
 $\pm$ 0.08 & 200$^c$\\
NGC 6888 (inner) & 54.62 $\pm$ 0.94 & 108.70 $\pm$ 1.85 & 4.28 $\pm$ 0.30 & -- & 4.06 $\pm$ 0.39 & 1.25
 $\pm$ 0.07 & 100$^c$\\
A 48             & 602.92 $\pm$ 0.52 & 680.93 $\pm$ 0.96 & 8.41 $\pm$ 0.09 & 0.18 $\pm$ 0.05 & 7.29 $\pm$ 0.81 & 0.49
 $\pm$ 0.06 & 700$^d$\\
\hline
	\end{tabular}
	\end{center}	
	
	$^a$: $\times$ 10$^{-16}$ W m$^{-2}$. $^b$: \citet{2011MNRAS.418.2532S}, $^c$: \citet{2012AA...541A.119F}, $^d$: \citet{2013arXiv1301.3994F}\\

\end{table*}

\subsection{Nebular Ionization Balance and Electron Densities}

As discussed by \citet{2001MNRAS.323..343L}, for a wide range of nebular 
excitation classes measured N$^{2+}$/O$^{2+}$ ratios are expected to be 
close to the overall nebular N/O ratios, due to the similar ionization 
potentials of the first and second ion stages of nitrogen and oxygen. 
However, the [N~{\sc iii}] 57-\micron\ and [O~{\sc iii}] 88-\micron\ lines 
have different critical densities, so nebular electron density estimates 
are needed when using these lines to evaluate nebular N$^{2+}$/O$^{2+}$ 
ratios.

The [N~{\sc ii}] 122, 205 \micron\ emission lines were observed with the 
intention of measuring their relative fluxes, since their flux ratio is an 
electron density diagnostic. However, the weaker of the two lines, at 205 
\micron, falls in a poorly calibrated and noisy wavelength region of the 
PACS spectrometer's wavelength coverage and only two detections of this 
line having a signal to noise ratio greater than three were obtained. The 
122-\micron\ line was detected, quite strongly in some cases, in 
all of the nebular spectra except for that of WR 8.

At the low density limit (e.g. $\sim$1~cm$^{-3}$) the flux ratio of 
the [N~{\sc ii}] 122/205-$\mu$m lines is predicted to be 0.74, and the 
ratio approaches its high-density limit ratio of $\sim$10 for electron 
densities above 2000~cm$^{-3}$. As can be seen from 
Table~\ref{tab:intline}, for the two cases where detections of [N~{\sc 
ii}] 205 \micron\ where made, the 122/205 flux ratio is found to be 25 and 56. 
Since the optically determined electron densities for all of the 
nebulae are below 1000~cm$^{-3}$, we attribute this discrepancy to the 
problems in calibrating PACS spectra longwards of 190~$\mu$m due to the 
leak of shorter wavelength radiation that affects this spectral region 
\citep{2010A&A2...518L...2P}. Instead, we have adopted electron densities measured at optical 
wavelengths (Table~\ref{tab:intline}). A number of different regions of 
NGC~6888 were studied by \citet{2012AA...541A.119F} and values of n$_e$ 
of 100 and 200 cm$^{-3}$ were found for fields close to those we observed 
with PACS. For WR~8, \citet{2011MNRAS.418.2532S} found a nebular electron 
density of n$_e$ $\sim$ 50 cm$^{-3}$, while for Abell~48 
\citet{2013arXiv1301.3994F} found an electron density of 700~cm$^{-3}$ 
- consistent with our strong 57 and 88 \micron\ line detections.

We have used our detections of the [N~{\sc ii}] 122-\micron\ and [N~{\sc 
iii} 57-$\mu$m lines to calculate the N$^{2+}$/N$^{+}$ abundance ratio for 
each of the three nebulae for which the [N~{\sc ii}] 122-\micron\ line was 
detected. The {\sc equib} code, developed at UCL by I. D. Howarth and S. 
Adams, was used to solve the statistical equilibrium equations for 
multilevel atomic models of N$^{2+}$ and N$^{+}$, giving level populations 
and line emissivities for the appropriate physical conditions. The atomic 
data used by {\sc equib} are from the {\sc chianti} atomic database 
\citep{2006ApJS..162..261L}. Detailed references for the collision strengths and transition probabilities can be found in Table~\ref{reftable}.

\begin{table}
	\centering
	\caption{Sources of Atomic data}
	\begin{tabular}{p{1.5cm}|p{6cm}}
	Ion &  References\\
	\hline
	N$^+$		&  A Values: \citet{1995PhyS...52..240B, 2000MNRAS.312..813S}; Collision Strengths: \citet{1994MNRAS.268..816S} \\
	N$^{2+}$	&  A Values: \citet{1979AA....71L...5N}; Collision Strengths: \citet{1994MNRAS.266..715S} \\
	O$^{2+}$	&  A Values: \citet{2000MNRAS.312..813S, doi10.1139p01-059}; Collision Strengths: \citet{1994AAS..103..273L} \\
	\end{tabular}
 \label{reftable}	
\end{table}

From the atomic data, the relationship between the 
line intensities and the ionic abundances was found. At the low density 
limit for both lines this resolves to N$^{2+}$/N$^{+}$ $\simeq$ 0.22 
$\times$ (I$_{57}$ / I$_{122}$), where I$_{57}$, I$_{122}$ are the fluxes 
of the 122- and 57-$\mu$m lines respectively. At the high density limit 
for both lines the factor of 0.22 becomes 0.08. For each nebula we 
calculated the appropriate factors using the literature electron densities 
listed in Table~\ref{tab:intline}). Conservative uncertainties of 50\% 
have been assumed for the electron densities and these are included in the 
final uncertainties quoted in Table~\ref{tab:intline}. For two of the 
three observed fields for which the [N~{\sc ii}] 122 \micron\ line was 
detected, the N$^{2+}$ / N$^{+}$ ratio is greater than unity, indicating that 
N$^{2+}$ is the dominant ionization state of nitrogen. Given the 
similarity of their ionization potentials, we conclude that O$^{2+}$ is also 
the dominant ionization state for oxygen in the observed fields.

\subsection{N/O Ratios and Nitrogen Enrichment}

For each nebula, the conversion factor to be divided into the measured [N~{\sc iii}] 
to [O~{\sc iii}] line intensity ratio to obtain the N$^{2+}$/O$^{2+}$ ratio was 
calculated using {\sc equib}. These factors ranged from around 1.5 at the 
lowest electron densities considered (50 cm$^{-3}$) to 2.3 at the highest 
density (700 cm$^{-3}$). The uncertainties quoted in 
Table~\ref{tab:intline} include an 50\% uncertainty in the electron 
densities adopted from the literature. From a comparison of the nebular 
N/O ratios found in Table~\ref{tab:intline} with the solar and Galactic 
H~{\sc ii} region N/O ratios of 0.12-0.14 (\citealt{2009ARA&A..47..481A}, 
\citealt{2011MNRAS.418.2532S}), it can be seen that the N/O ratios of the 
two massive WR nebulae are enriched by factors of 7 to 10 relative to the 
solar ratio, while the N/O ratio of the [WN] planetary nebula Abell~48 is 
enhanced by a factor of 4.

Our measured N/O ratios are consistent with previous measurements, which, 
due to the absence of lines of [N~{\sc iii}] in the optical region, had 
been based on the ratio of the ion stages N$^+$ and O$^+$. To provide 
context to the abundances calculated for our sample, we have included 
Table~\ref{abundtable}, which shows the results of prior studies, sorted by the
spectral type of the host star.

\begin{table*}
	\centering
	\caption{Enrichment of Nitrogen in nebulae around massive WR stars}
	\begin{tabular}{ccccp{4.5cm}}
	Nebula & Star & Spectral Type & Nebular N/O &  Reference\\
	\hline
	S 308    & WR 6   & WN4     & 0.54       & \citet{K84}  \\
	S 308    & WR 6   & WN4     & 1.65       & \citet{1992AA...259..629E}  \\
	NGC 2359 & WR 7   & WN4     & 0.11       & \citet{1992AA...259..629E}  \\	
	NGC 3199 & WR 18  & WN4     & 0.20       & \citet{K84} \\
	NGC 3199 & WR 18  & WN4     & 0.12       & \citet{2011MNRAS.418.2532S} \\
	RCW 104  & WR 75  & WN6     & 0.74       & \citet{1992AA...259..629E}  \\ 
	NGC 6888 & WR 136 & WN6h    & 0.50        & \citet{E92-6888} \\
	NGC 6888 & WR 136 & WN6h    & 0.50        & \citet{M00-6888} \\
	NGC 6888 (rim$^a$)   & WR 136 & WN6h    & 1.20       & \citet{2012AA...541A.119F} \\
	NGC 6888 (inner$^b$) & WR 136 & WN6h    & 0.97       & \citet{2012AA...541A.119F} \\
	NGC 6888 (inner) & WR 136 & WN6h    & 1.65       & \citet{2014arXiv1402.6181M} \\
	NGC 6888 (rim)   & WR 136 & WN6h    & 0.93       & this work \\
	NGC 6888 (inner) & WR 136 & WN6h    & 1.25       & this work \\
	anon     & WR 8   & WC4/WN7 & $>$ 0.4    & \citet{2011MNRAS.418.2532S} \\		
	anon     & WR 8   & WC4/WN7 & 0.92       & this work \\
	anon     & WR 16  & WN8h    & high       & \citet{M99} \\
	anon     & WR 16  & WN8h    & $>$ 0.8    & \citet{2011MNRAS.418.2532S} \\
	RCW 58   & WR 40  & WN8h    & 0.38       & \citet{K84} \\
	RCW 58   & WR 40  & WN8h    & $>$ 1.25   & \citet{2011MNRAS.418.2532S} \\
	M 1-67   & WR 124 & WN8h    & 2.95       & \citet{E91}  \\
	M 1-67   & WR 124 & WN8h    & 2.81       & \citet{2013AA...554A.104F} \\ 
	\end{tabular}
 \label{abundtable}
	
\bigskip	
	$^a$: Region X1, $^b$: region MB1.
\end{table*}

\subsubsection{WR~8} 

From optical spectroscopy of the nebula around WR~8, 
\citet{2011MNRAS.418.2532S} found N/O ratios in the range 
0.4~$<$~N/O~$<$~1.1, consistent with our measurement here of 0.92 $\pm$ 
0.17. Our non-detection of the far-infrared [N~{\sc ii}] lines is likely 
due to a combination of a low fractional N$^+$ abundance and the intrinsic 
faintness of the source. The nebula around WR~8 is likely to be composed 
of stellar ejecta processed by the CNO-cycle.  The enhanced nitrogen abundance 
is characteristic of material ejected during the red supergiant phase (see 
e.g. NGC 6888). The WN/WC type of the central star indicates that it is 
more evolved than the other known WR ejecta nebulae \citep{C95}, 
consistent with its fainter nebula. 

In comparison with the abundances found for other WR nebulae presented in Table~\ref{abundtable}, the 
N/O ratio we find for the nebula around WR 8 is actually rather modest. This, combined with the advanced 
evolutionary state of the system, suggests that there is a substantial component of swept-up ISM in the 
nebula. 

\subsubsection{NGC 6888}

Optical N$^+$/O$^+$ ratios were measured by \citet{2012AA...541A.119F} 
for NGC~6888, who found values of 0.74-1.20 near to the `rim' region for 
which we found an N${^{2+}}$/O$^{2+}$ ratio of 0.93 $\pm$ 0.08, and a 
N$^+$/O$^+$ ratio of 0.18-0.48 for a central region close to where we 
found N${^{2+}}$/O$^{2+}$ = 1.25 $\pm$ 0.07. The earlier optical 
spectroscopic study of \citet{E92-6888} found an N$^+$/O$^+$ ratio of 
$\sim$2 for the blue- and red-shifted shell components of the nebula, 
while an ambient gas velocity component showed a more solar-like 
N$^+$/O$^+$ ratio. A more recent study by \citet{2014arXiv1402.6181M} 
found a much higher value of N/O = 1.65 for a field closer to the exciting star 
than our inner field, possibly representing more recent stellar ejecta with respect to 
that measured at greater distances.

Clear differences are evident between the two NGC~6888 fields. The rim 
field has stronger [N~{\sc ii}] lines and an N$^{2+}$/N$^{+}$ ratio of 0.7, 
consistent with representing material far from the exciting star. 
In contrast, the central field shows much weaker [N~{\sc ii}] lines 
relative to the [N~{\sc iii}] 57-$\mu$m line.

The N/O ratios found for NGC 6888 are again moderate compared to the 
extreme examples in Table~\ref{abundtable}. The lower abundances in the rim 
of NGC 6888 suggest that the rim regions include a large swept-up ISM component. 
However, \citet{2012AA...541A.119F} found the opposite trend in that their highest 
N/O ratio was found for a bright section of the rim of NGC 6888. This could be an 
ionization effect though as the N/O ratio quoted by \citet{2012AA...541A.119F} is 
N$^+$/O$+$, and these ionization stages are unlikely to be dominant as discussed 
earlier.

\subsubsection{Abell 48} 

Abell~48 has been shown to be a planetary nebula with an early WN central star 
(\citealt{2013MNRAS.430.2302T2}, \citealt{2013arXiv1301.3994F}).
The former authors found an N$^+$/O$^+$ ratio of 0.25$\pm$0.10 for Abell~48, 
while the latter authors found an N$^+$/O$^+$ ratio of 0.12. In addition, an N/O ratio of 0.27 has been
quoted by \citet{2014MNRAS.tmp..389D}. Our N$^{2+}$- and O$^{2+}$-based measurement of N/O = 0.49 $\pm$ 0.06 does not agree with these measurements by almost a factor of two. However, the detailed photoionization models of \citet{2014MNRAS.tmp..389D} predict N$^{2+}$- and O$^{2+}$ abundances which, when combined yield an N$^{2+}$/O$^{2+}$ ratio of 0.52, which exactly match that observed.

Abell~48 has a large N$^{2+}$/N$^{+}$ ratio of 7.3 (Table~\ref{tab:intline}). 
The nebular spectra of \citet{2013MNRAS.430.2302T2} and 
\citet{2013arXiv1301.3994F} showed no He~{\sc ii} or [Ar~{\sc iv}] line 
emission, indicating a lack of stellar photons with energies in excess of 
54.4~eV, so we conclude that N$^{2+}$ and O$^{2+}$ are the dominant ion 
stages of nitrogen and oxygen and that our N$^{2+}$/O$^{2+}$-based N/O 
ratio for Abell~48 should therefore be accurate.

The other three objects in the rare class of planetary nebulae having WN 
central stars have been found to have N/O ratios ranging from 0.28 to 0.4 
(LMC-N66: \citealt{2003MNRAS.345..186T}, N/O = 0.4; PB 8: 
\citealt{2009A&A...496..139G}, N/O = 0.28; IC 4663: \citealt{2012MNRAS.423..934M}, N/O = 0.36). 
Each of these studies had access not only to the optical lines of [N~{\sc 
ii}] but also to the ultraviolet N~{\sc iii}] and N~{\sc iv}] lines. Our 
measurement of N/O = 0.49 $\pm$ 0.06 for Abell~48 falls slightly above the
N/O ratios found for the Galactic members of this class (IC 4663 and PB 8).

\subsection{Predicted Stellar Yields and Swept-Up material}

\citet{2011ApJ...737..100T} computed models for the ejecta from massive 
stars and calculated the degree of nitrogen enrichment that would be 
expected in the nebulae as a function of time. They found that the N/O 
values seen in NGC~6888 and S~308, the two best known WR bubbles, are 
characteristic of a $\sim$ 3 $\times$ 10$^5$ year WR period in the life of 
40-60 M$_{\sun}$ stars (see Figure~16, \citealt{2011ApJ...737..100T}). 
They are also consistent with the N/O ratios that we find here for
NGC~6888 and WR~8.

However, it should be noted that the results of \citet{2011ApJ...737..100T} 
do not include the effect of the nebulae sweeping up ISM material and hence diluting 
the detected nebular abundances. From Table~\ref{abundtable} it is clear that the 
highest N/O ratios detected for classical WR nebulae are for M 1-67, one of the most 
compact of the class with a radius of around 1.2 pc \citep{2013AA...554A.104F}. 
If we assume that this is a reasonable benchmark for the abundances of a WR nebula
containing very little swept up material, it is clear that the other nebulae listed 
in Table~\ref{abundtable} must suffer from considerable dilution of their natal 
abundances. For consistency then, models such as those of \citet{2011ApJ...737..100T} 
should find much higher N/O ratios than observed, such that they would reproduce the 
observed abundances after ISM dilution was taken into account.

\section{Conclusions}

The degree of nitrogen enrichment in a selection of nebulae around WR 
stars has been measured using Herschel/PACS observations of their far-IR fine 
structure lines. Abundance studies are usually performed using optical 
spectra, which has the drawback of not including any lines of N$^{2+}$, 
which is often the dominant ionization stage for nitrogen. This effect is 
usually compensated for using ionization correction factor schemes to 
calculate the N$^{2+}$ abundance. The far-IR observations used in this 
paper include the 57 \micron\ line of N$^{2+}$, which should result in 
nitrogen abundances that are less susceptible to systematic errors. In addition, 
abundances obtained for N$^+$, N$^{2+}$ and O$^{2+}$ from their 
far-infrared lines have little or no sensitivity to the nebular electron 
temperature, due to their low excitation energies.

The N/O ratios of 0.9 -- 1.3 found for the massive WR-star nebulae WR~8 and 
NGC 6888 are consistent with both nebulae representing material ejected 
during the pre-WR evolution of the stars. For NGC 6888, our measurements 
overlap those of previous authors who employed ionization correction 
factors to determine the N/O ratio from optical lines. Our observations 
confirm that the nebula around WR~8 has a significant component of ejected 
stellar material -- as predicted by the morphological categorization 
methods of \citet{1991IAUS..143..349C} which were employed by 
\citet{2010MNRAS.409.1429S} upon discovery of the nebulosity.

The other object observed, Abell~48, is now known to be a rare type of planetary nebula
with a WN4-type central star {and which are not of Peimbert Type I. Our far-infrared observations yield
a nebular N/O ratio of 0.49, which is slightly greater than the ratios found for the two  
other Galactic members of the rare [WN] class of planetary nebulae.

\section*{Acknowledgments}

We thank the anonymous referee for their suggestions which have helped improve and clarify the paper. DJS acknowledges support from an NSERC Discovery Grant and an NSERC Discovery Accelerator Grant. This research has made use of NASA's Astrophysics Data System Bibliographic Services.

DJS thanks A. Danehkar for alerting us to a mistake in the unit conversions in the first version of this paper.


\bibliographystyle{mn2e}

\label{lastpage}

\end{document}